\documentclass[draft,grl]{agutex}
\usepackage{graphicx,color,ulem,lineno}

\setkeys{Gin}{draft=false}

\authorrunninghead{Y\.I\u G\.IT ET AL.}
\titlerunninghead{GRAVITY WAVES ON MARS}

\authoraddr{Corresponding author: E. Yi\u git, George Mason University, 
Department of Physics and Astronomy, Space Weather Group, 
MSN:3F3, Fairfax, VA 22030, USA. (eyigit@gmu.edu)}

\graphicspath{{/Users/erdal/Research/MAVEN/Figures/}}

\begin{document}
\title{High-altitude gravity waves in the Martian thermosphere observed by
  MAVEN/NGIMS and modeled by a gravity wave scheme} 

\authors{Erdal Yi\u git\altaffilmark{1}, Scott L. England\altaffilmark{2},
  Guiping Liu \altaffilmark{2}, Alexander S. Medvedev\altaffilmark{3,4}, Paul
  R. Mahaffy\altaffilmark{5}, Takeshi Kuroda\altaffilmark{6}, 
  Bruce M. Jakosky\altaffilmark{7} }

\altaffiltext{1}{George Mason University, Department of Physics and Astronomy,
  Fairfax, VA, USA.}  \altaffiltext{2}{University of California, Berkeley, Space
  Sciences Laboratory, Berkeley, CA, USA.}  \altaffiltext{3}{Max Planck
  Institute for Solar System Research, G\"ottingen, Germany.}
\altaffiltext{4}{Institute of Astrophysics, Georg-August University,
  G\"ottingen, Germany.}  \altaffiltext{5}{NASA, Goddard Space Flight Center,
  Greenbelt, MA, USA.}  \altaffiltext{6}{Department of Geophysics, Tohoku
  University, Sendai, Japan.} \altaffiltext{7}{Laboratory for Atmospheric and Space Physics,
University of Colorado, USA.} 

\newpage
\begin{abstract}
  First high-altitude observations of gravity wave (GW)-induced CO$_2$
  density perturbations in the Martian thermosphere retrieved from NASA's NGIMS
  instrument on board the MAVEN satellite are presented and interpreted using
  the extended GW parameterization of \citet{Yigit_etal08} and the Mars Climate
  Database as an input. Observed relative density perturbations
  between 180--220 km of 20--40\% demonstrate appreciable local time, latitude,
  and altitude variations. Modeling for
  the spatiotemporal conditions of the MAVEN observations suggests that
  GWs can directly propagate from the lower atmosphere to the thermosphere,
  produce appreciable dynamical effects, and likely contribute to the observed
  fluctuations. Modeled effects are somewhat smaller than the observed but their
  highly variable nature is in qualitative agreement with observations. Possible
  reasons for discrepancies between modeling and measurements are discussed.
\end{abstract}

\begin{article}
\section{Introduction}
\label{sec:intro}
Internal gravity waves (GWs) are fundamental features of all stably stratified
planetary atmospheres. On Earth, they are primarily generated by lower
atmospheric meteorological processes, and have profound dynamical and thermal
effects on the circulation of the middle and upper atmosphere
\citep[e.g.,][]{Yigit_etal09}. A contemporary review of the role of internal
wave processes in Earth's atmosphere can be found in the work by
\citet{YigitMedvedev15} and in the book by \citet[][Chapter 5]{Yigit15}. On
Mars, GWs of lower atmospheric origin play a similar important dynamical
\citep{Medvedev_etal11b} and thermal role \citep{MedvedevYigit12} in the upper
atmosphere.  While global modeling demonstrates a direct influence of GWs on the
large-scale Martian circulation, observations have provided a valuable insight
into various GW structures. Although the effect of waves cannot be directly
measured, GW signatures, for example, small-scale density perturbations have
been detected during aerobraking on Mars Global Surveyor and Mars Odyssey
\citep[e.g.,][]{Fritts_etal06}. Such measurements have shown that magnitudes of
GW-induced perturbations vary between 5--50\%.

NASA's Mars Atmosphere Volatile EvolutioN (MAVEN) mission was launched in
November 2013 and entered Mars' orbit in September 2014. Its prime goal is to
explore the Martian upper atmosphere in an unprecedented manner and to help
better understand Mars atmospheric loss processes and coupling to the solar
wind. In this study, we analyze the first CO$_2$ density data from the Neutral
Gas Ion Mass Spectrometer (NGIMS) instrument on board the MAVEN satellite for
high-altitude gravity wave signatures, and interpret these results by
calculating for the spatiotemporal conditions of the observations the
fluctuations associated with the propagation of lower atmospheric GWs into the
thermosphere and the direct effects resulting from their dissipation.  For this,
the GW parameterization of \citet{Yigit_etal08} is used together with the Mars
Climate Database (MCD) atmospheric fields.

Next section presents the MAVEN data and the analysis method; section
\ref{sec:tools} outlines the GW scheme that calculates the direct GW propagation
in the Martian whole atmosphere; sections \ref{sec:maven-obs} and
\ref{sec:gravity-wave-induced} present the observed density fluctuations and the
modeled GW effects and fluctuations, respectively. Summary and conclusions are
given in section \ref{sec:conc}.

\section{MAVEN/NGIMS Data \& Analysis Method}\label{sec:maven-data}

The NGIMS instrument on board the MAVEN satellite is a quadrupole mass
spectrometer designed to measure the density of the gas in Mars' upper
atmosphere between 2 and 150 Daltons with unit mass resolution
\citep{Mahaffy_etal14}. It has an open and a closed source channel, both of
which are used to measure the neutral density when MAVEN is below 500 km
altitude, relative to the areoid. Both channels can be used to measure the
density of CO$_2$, which is a useful proxy for the total mass density at
altitudes below $\sim$250 km.

This study uses the Level 2 (version 3, revision 1) of the NGIMS data, in which
the instrument counts in each mass channel have been combined, knowing the
fractionation patterns of CO$_2$, into a mass density for this species.
Further, the effect of the transition from the unattenuated instrument counts at
high altitudes to the attenuated instrument counts at lower altitudes has been
accounted for, using data from an altitude when both the attenuated and
unattenuated signals are within the appropriate range.  The analysis presented
here will focus on determining atmospheric density perturbations seen in either
one of open or close source channels.  As such, any uncertainties in the
differences between these channels, how densities are computed from either
channel, any uncertainty in the fractionation of CO$_2$, or any inbound/outbound
asymmetry resulting from the pile-up effect do not affect the analysis presented
here. Given that the NGIMS instrument sweeps across all mass channels with a 4 s
cadence, it can be used to identify perturbations with spatial scales larger
than $\sim$20 km along the spacecraft track.  With an orbital speed of $\sim
4.2$ km s$^{-1}$ the geometry of this sampling varies during the 12 minutes that
the spacecraft spends at altitudes below 250 km.  Close to periapsis, the
spacecraft is moving almost entirely parallel to the areoid, and thus would be
sampling only horizontal structures (wavelengths) as it moves, whereas at higher
spacecraft altitudes there is an increasing vertical component to its motion,
and as such vertical structures (wavelengths) will become increasingly important
in any signal seen by this instrument.

Figure 1a shows the CO$_2$ density measurements from a typical periapsis pass.
For this particular orbit, periapsis was at 153 km above the areoid, at
74$^\circ$N, 179$^\circ$E, 4.2 hours local time and 259$^\circ$ solar longitude.
The total time that the spacecraft is below 250 km above the areoid is 710 s.
Examining the overall structure of the CO$_2$ density profile during this
periapsis pass, a clear increase in CO$_2$ density with decreasing spacecraft
altitude can be seen.  However, superimposed on this trend are significant,
quasi-oscillatory perturbations in the density, whose amplitudes also increase
towards periapsis.  These
perturbations are visible at spacecraft altitudes from periapsis (153 km) up to
around 200 km, which is several scale heights above the altitude where GWs have 
been seen in previous thermospheric observations at Mars. For
example, \citet{Tolson_etal07} had reported wave observations between
$\sim$100--140 km altitude.

\begin{figure}
  \centering
  \includegraphics[clip,width=1\columnwidth]{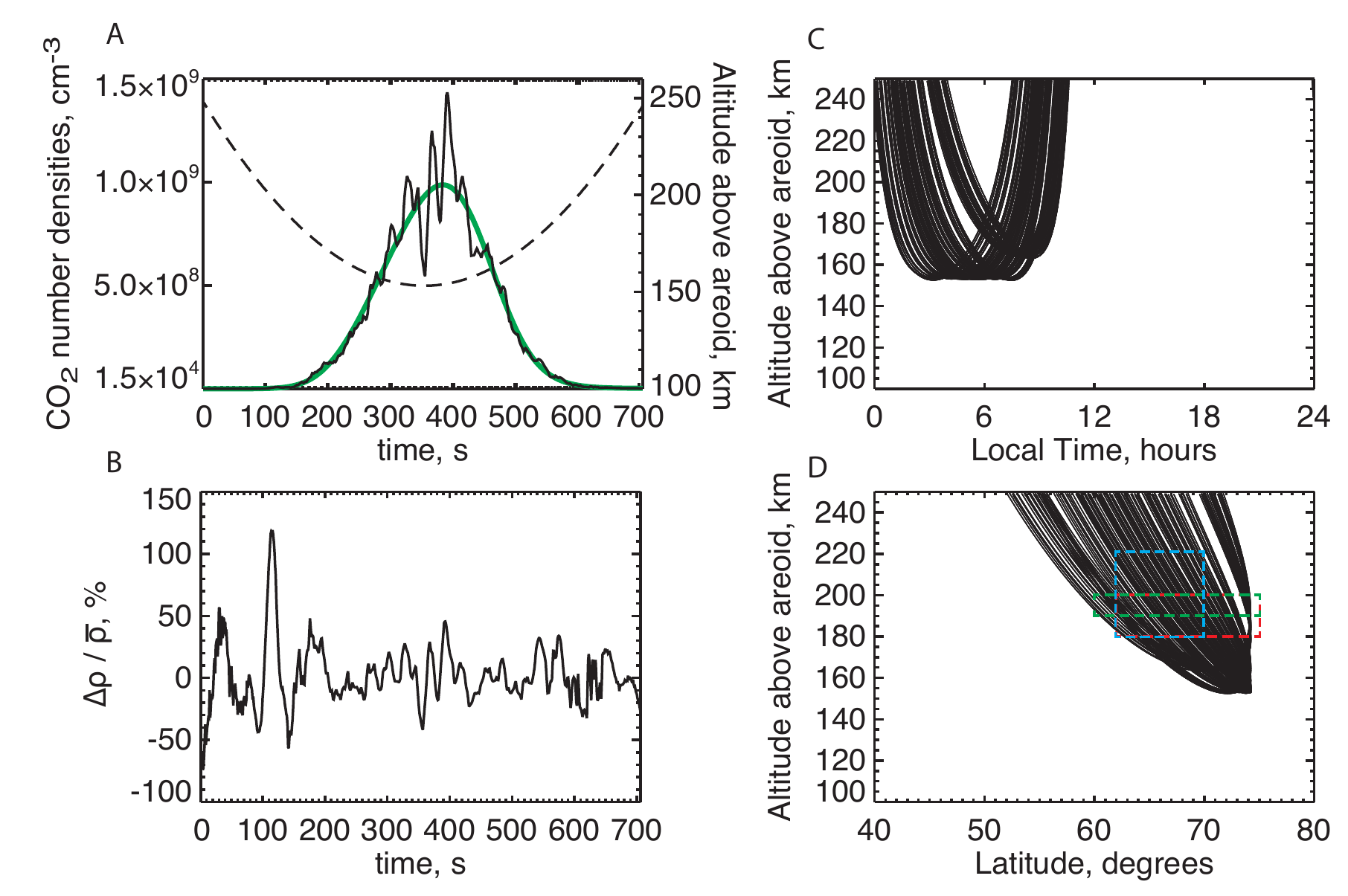}
  \vspace{-0.1cm}
  \caption{MAVEN orbital details: (a) The black dashed line shows the MAVEN
    spacecraft altitude as a function of time measured from the point at which
    the spacecraft passes 250 km altitude relative to the areoid on its inbound
    pass (18 December 2014, 18:41:04 UT).  The solid black line shows the
    CO$_2$ number density measured by NGIMS during this periapsis pass. The
    solid green line shows the least-squares fit to the CO$_2$ number density
    described in the text; (b) shows the relative density perturbations during
    the same periapsis pass shown in panel (a);  (c) and (d) show the sampling
    of the NGIMS data as functions of altitude, local time and latitude, for all
    the data from December 2014 that are used in this study, where each dot
    represents a single CO$_2$ density measurement.  The red, blue, and green
    boxes are described in the text.}
  \label{fig:fig1}
\end{figure}

To identify the perturbations, an estimate of the background density profile is
necessary. Previous analysis of similar in situ observations have done this
using a variety of techniques including a long running mean
\citep[e.~g.,][]{Kasprzak_etal88},
and a least-squares polynomial fit \citep[e.~g.,][]{Snowden_etal13}.  The green
line on Figure 1a shows a 7th order polynomial fit to the log of the density
observations, which provides a good estimate of the background mean density and
allows for (1) fitting the overall density increase towards periapsis, (2)
changes in the scale height (and therefore temperature) of the background over
the altitude range fitted, and (3) for asymmetry between the inbound and
outbound density measurements resulting from differences in the sampling
latitude, longitude and local time between the inbound and outbound portions of
the periapsis pass and any signature instrumental sampling artifacts.  Removing
this background from the observations provides an estimate of the density
perturbation, which is then normalized against the background density to provide
an estimate of the fractional density perturbation ($\Delta \rho /
\bar{\rho}=\rho^\prime/\rho_0$).  This is shown for the same periapsis pass in
Figure \ref{fig:fig1}b. The associated density perturbations are
quasi-oscillatory, having a range of amplitudes of order $10-100\%$ of the
background mean density and exist throughout the entire periapsis pass, although
both the count rate of the instrument and the ability to fit to the background
density decrease with increasing altitude.  For this reason, we will limit the
analysis of these perturbations to observations taken below 220 km altitude
($\sim$50--650 s during this periapsis pass), and focus primarily on
observations taken below 200 km altitude ($\sim$100--600 s during this periapsis
pass). Further, as these perturbations are quasi-oscillatory, with spatial
scales of tens to hundreds km along the satellite track, the assumption that
they are associated with atmospheric GWs will be made in all subsequent
analysis, following previous studies \citep[e.g.,][]{Kasprzak_etal88,
  Fritts_etal06, Tolson_etal07}. While there is likely some component of the
observed density perturbation not associated with such waves, if GWs dominate
these observations, the mean behavior of the density perturbations is expected
to follow that of the atmospheric GWs in this region.  For this reason, all
subsequent analysis will determine the mean properties of these density
perturbations over a region of the Martian thermosphere.

This study focuses on data taken during December 2014. During 
this time period, there was a latitudinally-symmetric coverage in the
middle-to-high Northern latitudes as the MAVEN spacecraft moved from day to
night across the dawn terminator during the periapsis passes. This allowed us to
look for day-to-night variations in GWs, which happened to be one of the key
 objectives from the Venus Express Orbiter Neutral Mass
Spectrometer (ONMS) data \citep[e.g.,][]{Kasprzak_etal88}. Finally, the
periapsis altitude remained approximately constant during this time period,
which enables the averaging of the data from this entire time interval to
compute the mean amplitude of the observed GWs and their 
variation.  In total, there are 78 periapsis data sets for which good quality
Level 2 densities are available during the time period covering 1--23 December
for the closed source data, and 1--18 December for the open source data. The
altitude, local time and latitude coverage of these data is shown in Figures
\ref{fig:fig1}c and \ref{fig:fig1}d.  To calculate the mean properties of the
observed GWs, the region over which data were averaged was chosen based on the
coverage shown in Figures \ref{fig:fig1}c and \ref{fig:fig1}d to avoid any
sampling bias, such as the correlation between altitude and latitude which would
occur if the entire data set were used.

\section{The Extended Gravity Wave Scheme}\label{sec:tools}
The extended spectral nonlinear gravity wave parameterization of
\citet{Yigit_etal08} is employed here for quantifying the direct propagation of
lower atmospheric GWs into the Martian thermosphere. Mars Climate Database
fields (meridional and zonal wind velocities, density, and temperature)
appropriate for the spatiotemporal conditions of the presented MAVEN
observations are used as an input in order to drive the extended
parameterization and calculate the GW-induced zonal gravity wave drag as well as
wind and density fluctuations.

The extended GW parameterization, developed originally for Earth's whole
atmosphere system models, is described in detail in the work by
\citet{Yigit_etal08}. Its implementation and application to modeling Mars'
atmosphere is given in the papers by \citet{Medvedev_etal11a, Medvedev_etal11b}.
The scheme is appropriate for planetary whole atmosphere numerical models
extending from the lower atmosphere into the thermosphere. Its extensive use in
general circulation modeling of the terrestrial \citep{Yigit_etal09,
  Yigit_etal12b, Yigit_etal14, YigitMedvedev09, YigitMedvedev10} and Martian
\citep{MedvedevYigit12, Medvedev_etal13, Medvedev_etal15, Yigit_etal15}
atmosphere demonstrated an appreciable vertical coupling by gravity waves in
both planets, and have thus helped to quantify wave-mean flow interactions
there. Application of this scheme has allowed a Martian general circulation
model to reproduce for the first time the latitudinal temperature profile
derived from the aerobraking measurements of \citet{Bougher_etal06}.

The gravity wave scheme calculates vertical propagation and evolution of
small-scale wave harmonics systematically accounting for the major wave
dissipation mechanisms in the Martian atmosphere: due to nonlinear
breaking/saturation, molecular viscosity and thermal conduction. An empirical
distribution of GW spectrum is specified at the source level in the lower
atmosphere with a GW horizontal wavelength of $\lambda_H=250$ km and a source
strength of $\overline{u^\prime w^\prime}_{max} = 0.0025$ m$^2$ s$^{-2}$, as
described by \citet{Medvedev_etal11a}, and the scheme calculates at consecutive
vertical levels the momentum and temperature tendencies imposed by GWs on the
larger-scale atmospheric flow. In this study, we use the same GW source
specification (constrained with the radio occultation data of
\citet{Creasey_etal06a}) as in the previous works \citep{MedvedevYigit12,
  Medvedev_etal13}. Therefore, our modeling framework can provide a first-order
estimate of how much lower atmospheric GWs contribute to the observed
high-altitude density fluctuations. We next present the observations of GW
signatures by MAVEN, and then discuss the relevant modeling results.

\section{MAVEN Observations of CO$_2$ Density Fluctuations}\label{sec:maven-obs}

Despite the relatively constant observing geometry, substantial variations in
the gravity wave density perturbations along the spacecraft track are seen over
the 78 periapsis datasets taken during December 2014 as described in
Section~\ref{sec:maven-data} and shown in Figure~\ref{fig:fig1}a.  
GW signatures can be highly localized and variable
\citep[e.g.,][]{Creasey_etal06b, Fritts_etal06}; additionally the in situ
sampling of the NGIMS instrument is of local nature. Thus, rather than focusing
on the samples seen in individual orbits, which are subject to significant
geometric effects, the following analysis will focus on the mean properties
observed within a volume.  With the number of samples available during December
2014, this allows the variations in gravity wave amplitudes with latitude, local
time and altitude to be determined over a region that has never been studied
before at Mars.

\begin{figure}
  \centering
  \includegraphics[clip,width=1\columnwidth]{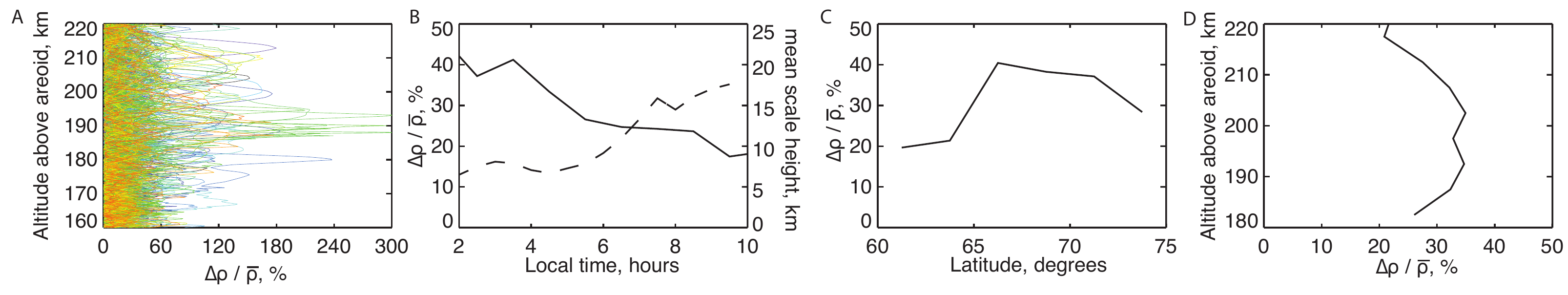}
  \vspace{-0.1cm}
  \caption{MAVEN observations of gravity wave-induced CO$_2$ density
    fluctuations in the upper thermosphere. (a) The relative density
    perturbations for all 78 profiles each displayed with different colors (both
    inbound \& outbound); (b) the mean value of the absolute relative density
    perturbation (solid line) and the mean scale height (dashed line) as a
    function of local time, for within 180--200 km altitude above the areoid and
    $62^\circ-75^\circ$N, from all periapsis passes during December 2014
    described in the text. Values are shown in one-hour local time bins; (c) as
    (b), but as a function of latitude, for all data within 190--200 km altitude
    above the areoid and $60^\circ-75^\circ$N.  Values are shown in 2.5$^\circ$
    latitude bins; (d) as (a), but as a function of altitude above the areoid,
    for all data within $62^\circ-70^\circ$. Values are shown in 5 km altitude
    bins.}
  \label{fig:fig2}
\end{figure}

To determine the local-time variation in gravity wave amplitudes, all data from
180-200 km altitude, and from 62$^\circ$-75$^\circ$ latitude are selected (shown
by the red box in Figure \ref{fig:fig1}d). The upper limit of 200 km is selected
to stay within the altitude range of both high CO$_2$ counting rate (clear
signal of the perturbations), and within the region where the background density
profile can be fit well. The lower limit of 180 km is chosen to restrict the
data to a region of approximately constant spacecraft velocity vector relative
to the surface (i.e., away form periapsis where the spacecraft is sampling
horizontal variations only).  The latitude range is selected to allow for even
sampling in latitude and altitude (see Figure \ref{fig:fig1}d).  Figure
\ref{fig:fig2}b shows with the solid line the mean values of the absolute
relative density perturbation ($\rho^\prime/\rho_0$) within this volume, in
1-hour local time (LT) bins.  A clear increase in the wave amplitudes is seen
with decreasing LT from a value of $\sim$20\% at 9-10 hours LT, to ~40\% at 2--3
hours LT.  This factor of two increase from day to night is comparable to that
found with the ONMS mass spectrometer observations on Pioneer Venus
\citep{Kasprzak_etal88}. The continuity equation implies that the density
perturbations resulting from GWs are the result of a combination of vertical
advection and adiabatic expansion. \citet{DelGenio_etal79} have argued that for
heavy species such as CO$_{2}$, whose density decreases rapidly with altitude,
the effect of vertical advection dominates. This effect is inversely
proportional to the density scale height $H$. Figure~\ref{fig:fig2}b shows the
variation of the CO$_{2}$ scale height with local time (dashed line). The mean
CO$_{2}$ scale height on the dayside is approximately twice that of the
nightside. Thus, one would expect the relative density amplitudes on the dayside
to be around half that of the nightside, which is consistent with the NGIMS
observations.

To determine the latitudinal variation in GW
amplitudes, all data from 190--200 km altitude, and from 60$^\circ$--75$^\circ$N
latitude are selected (shown by the green box in Figure \ref{fig:fig1}d). This
range is chosen to maximize the latitude range, while eliminating any
correlation between altitude and latitude within the chosen volume.  Figure
\ref{fig:fig2}c shows the mean values of the absolute relative density
perturbation over this range in 2.5$^\circ$ latitude bins.  A simple trend in
the wave amplitudes with latitude is not seen, but the amplitudes are somewhat
lower in the 60$^\circ$-65$^\circ$ latitude region ($\sim$20\% relative density
perturbation) compared with the region above 65$^\circ$N latitude ($\sim$30-40\%
relative density perturbation). Given that a comparatively small range of
latitude is sampled, it is possible that a more systematic latitudinal variation
could be present.  However, owing to the precession of the MAVEN spacecraft
orbit, it is not possible to decouple the latitudinal and local time variations
in the observations over any other subsequent portion of its orbit so-far,
limiting this analysis to just December 2014 and thus this relatively small
latitude range.

To determine the altitude variation in GW amplitudes, all data from 180--220 km
altitude, and from 62$^\circ$-70$^\circ$ latitude are selected (shown by the
blue box in Figure \ref{fig:fig1}d).  This range is chosen to maximize the
altitude range, while eliminating any correlation between the altitude and
latitude within the chosen volume.  Figure \ref{fig:fig2}c shows the altitude
variation of the mean $\rho^\prime/\rho_0$ in 5 km altitude bins:
$\rho^\prime/\rho_0$ increases from $\sim$25\% to 35\% up to 200 km level,
around which $\rho^\prime/\rho_0$ is approximately constant. Above 200 km
$\rho^\prime/\rho_0$ decreases down to 20\%. These results suggest that for the
observed waves for $z<200$ km exponential amplitude growth dominates over
dissipation, while for $z>200$ km wave dissipating becomes more dominant,
leading to weaker relative fluctuations.

Overall, these results indicate that the thermosphere is characterized by strong
dissipation of gravity waves and deposition of their energy into the background
atmosphere. This result is consistent with observations from lower altitudes in
the thermosphere, such as \citet{Fritts_etal06}, who presented evidence for
strong gravity wave dissipation in the 100--140 km altitude region. Modeling
studies have demonstrated strong GW dissipation in the Martian lower
thermosphere \citep[e.g.,][]{Medvedev_etal11b,MedvedevYigit12}. Our results
suggest that such dissipative processes and the associated wave dynamical and
thermal impact on the global circulation can extend to much higher altitudes.

A longitudinal variation in the gravity wave amplitudes, similar to that
reported by \citet{Fritts_etal06}, was investigated using the December 2014
NGIMS dataset (not shown), but no clear trend was found. It is possible that
this is a result of the higher altitude of the NGIMS observations, the limited
number of samples, or the time-period and latitude selected.

\section{Modeling Gravity Wave-Induced Effects on Mars}
\label{sec:gravity-wave-induced}

To provide a further perspective to the interpretations of observed density
fluctuations, we next model the direct propagation of lower atmospheric GWs into
the Martian thermosphere with the extended nonlinear GW scheme of
\citet{Yigit_etal08}. Using the MCD atmospheric fields chosen for the MAVEN
spatiotemporal conditions as an input to the GW scheme, we assess the
contribution of lower atmospheric GWs to the observed density fluctuations.

\begin{figure}
  \centering
  \includegraphics[clip,width=1\columnwidth]{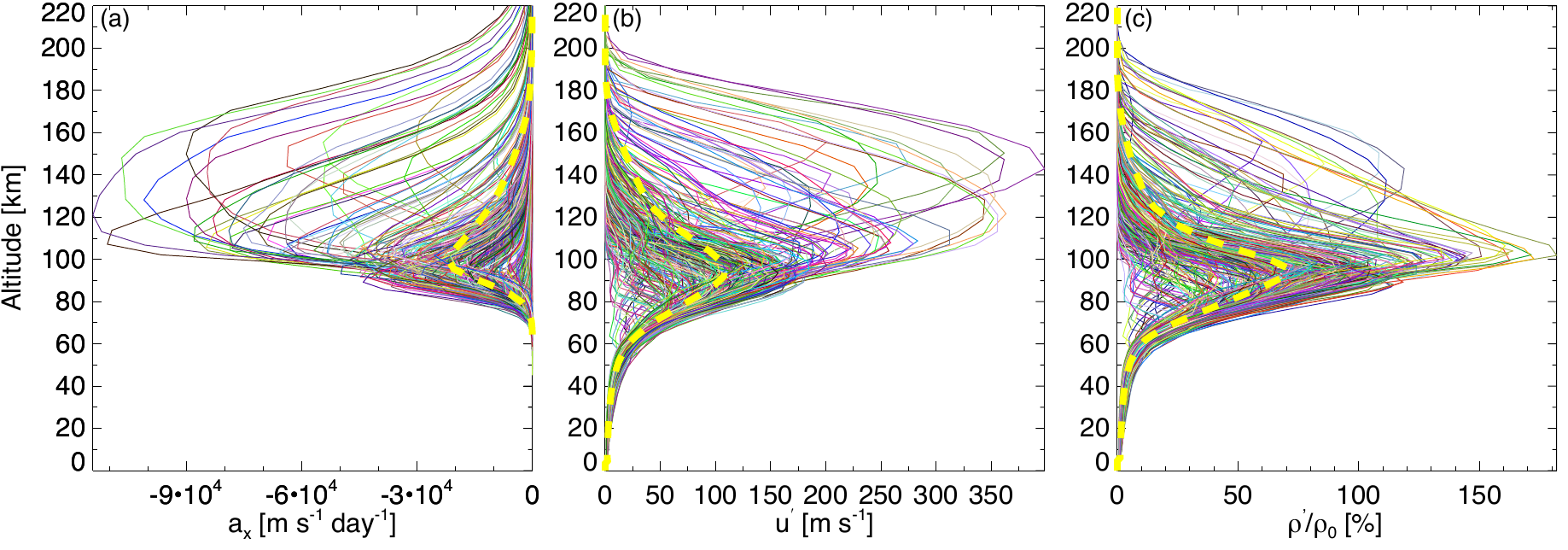}
  \vspace{-0.1cm}
  \caption{Altitude profiles of gravity wave effects calculated by the extended
    nonlinear gravity wave scheme of \citet{Yigit_etal08} using the Mars Climate
    Database atmospheric fields as input for the observational conditions of
    MAVEN. Each profile is shown with a different color: (a) zonal GW drag $a_x$
    (m s$^{-1}$ day$^{-1}$) with negative indicating westward drag; (b) rms wind
    fluctuations $u^\prime$ (m s$^{-1}$); (c) relative density perturbations
    $\rho^\prime/\rho_0$ (\%). The mean values of each parameter are shown with
    thick dashed yellow lines.}
  \label{fig:fig3}
\end{figure}

Figure \ref{fig:fig3} shows all the calculated instantaneous altitude profiles
of (a) the zonal GW drag $a_x$, (b) the root-mean square (rms) horizontal wind
fluctuations $u^\prime$, and (c) the relative density perturbations
$\rho^\prime/\rho_0$ modeled for MAVEN observational conditions. While the
modeled GW drag can be interpreted as representative of the wave dynamical
effects on the atmospheric circulation, $u^\prime$ provides an insight into
the propagation depth of small-scale GWs into the thermosphere. Overall, GWs
propagate to high altitudes (up to $\sim$200 km) and produce substantial drag of
up to $\sim$--10$^5$ m~s$^{-1}$~day$^{-1}$, peaking between 100 and 160 km. The
associated wind fluctuations, typically peaking around the same heights as well,
can reach up to 350 m~s$^{-1}$. The resulting relative density fluctuations can
reach instantaneously up to 180\% around 100 km. Overall, in regions of
strong GW activity (large $u^\prime$), wave-induced
acceleration/deceleration, which depends also on the rate of vertical wave
  damping, and density fluctuations are, generally, larger.
\citet{Fritts_etal06}'s analysis of MGS and ODY aerobraking density data
  showed significant (from tens to over 100 percent) GW-induced density
  fluctuations between 100 and 140 km.  As seen, our simulation results are in a
  very good agreement with the observations of \citet{Fritts_etal06} in this
  altitude range.

The average values of GW effects are also shown in Figure~\ref{fig:fig3} with
the  yellow dashed lines for each parameter. The peak mean
values are situated around 100 km with $a_x=-2000$ m~s$^{-1}$~day$^{-1}$,
u$^{\prime} = 100$ m~s$^{-1}$, and $\rho^{\prime}/\rho_0 \approx 70\%$. It is
seen that large departures can occur from the mean quantities.

These results demonstrate that GWs propagate very high, and their effects (as a
consequence of wave dissipation) are spatiotemporally very large. In the lower
atmosphere below $\sim 50$ km, GW dissipation is generally small, and individual
harmonics achieve large amplitudes as they arrive at higher altitudes, where
they are gradually obliterated by the increasing nonlinear damping, molecular
diffusion and thermal conduction \citep{Yigit_etal08,
  Medvedev_etal11b}. Overall, these results agree with the highly variable
nature of GW fluctuations observed by MAVEN/NGIMS. However, in the majority of
the calculated profiles, GWs happen to dissipate rapidly above $\sim$120 km:
75.7\% of the harmonics demonstrate decreasing amplitudes, suggesting that the
vast majority of the waves are being attenuated; only 5.7 \% of waves show
increasing amplitudes; and 18.6\% of harmonics have already been completely
dissipated below 120 km.

\begin{figure}
  \centering
  \includegraphics[clip,width=1\columnwidth]{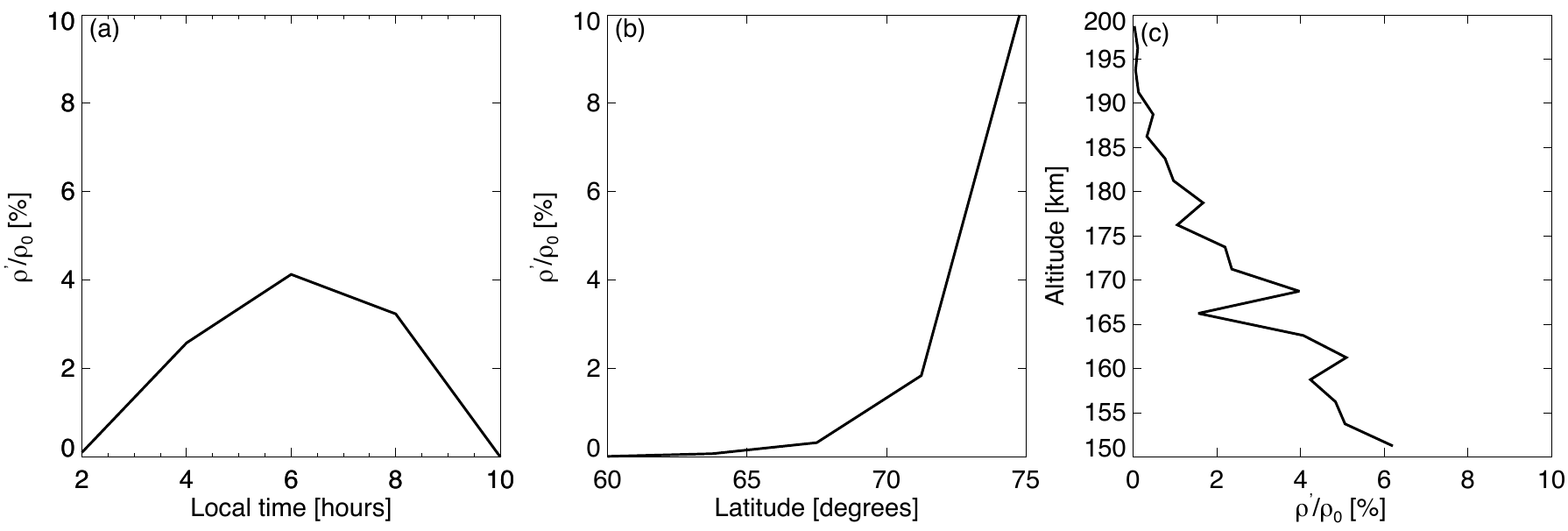}
  \vspace{-0.1cm}
  \caption{Modeled mean gravity wave-induced relative density perturbations
    $\rho^\prime/\rho_0$ (\%) that are presented in Figure \ref{fig:fig3} are
    binned for data between 150 and 200 km as a function of all (a) local times
    2--10 hours; (b) latitudes 60$^\circ$--75$^\circ$. Altitude variation
    between 150--200 km for all local times and latitudes are shown in panel
    (c).}
  \label{fig:fig4}
\end{figure}

Further data-model comparison can be performed by analyzing the spatiotemporal
variations of the modeled GW-induced density perturbations in the same manner as
was done for the NGIMS/MAVEN observations shown in Figure~\ref{fig:fig2}. Local
time, latitude, and altitude variations of $\rho^{\prime}/\rho_0$ are presented
in Figures~\ref{fig:fig4}a--c, respectively. In the calculations of the local
time (panel a) and latitude (panel b) variations of the mean
$\rho^{\prime}/\rho_0$, altitude levels from 150--200 km are considered. It is
seen that modeled values of $\rho^{\prime}/\rho_0$ are generally smaller than
those observed. GW-induced density perturbations peak at 6 LT, which partially
agrees with MAVEN. The latitude distribution shows the general tendency of GWs
to cause larger effects at higher latitudes.

There are several possible reasons for discrepancies between the simulated and
observed density fluctuations. One is related to the uncertainties of GW sources
in the lower atmosphere. Measurements of the GW field are difficult there due to
its quasi-chaotic behavior and very small wave amplitudes. Even on Earth, the
knowledge of GW sources is insufficient, and this is especially true for
Mars. Second, wave propagation is strongly affected by the background wind. It
was shown that the MCD fields, which are simply outputs from the general
circulation model, are obviously inconsistent with the GW feedback on the mean
flow \citep{Medvedev_etal11a}, and that the winds in the upper mesosphere and
thermosphere significantly alter when GW effects are included interactively in
simulations \citep{Medvedev_etal11b, Medvedev_etal13}: they decrease, and even
reverse, thus reducing the filtering and providing more favorable conditions for
GW penetration to higher altitudes. Third, the profiles measured by NGIMS are
the results of not only harmonics propagating from below and captured (to a
certain degree) by the parameterization. GWs generated at higher altitudes and
propagating significant horizontal distances are well known in the terrestrial
thermosphere.  Last but not least, the parameterization does not include fast
infrasound waves, which, as has been shown in the works by
\citet{WalterscheidHickey05} and \citet{Walterscheid_etal13}, can effectively
propagate to the upper thermosphere and attain significant amplitudes there.
Despite these limitations, the presented calculations provide a valuable insight
into the contribution to the detected variations of GWs of lower atmospheric
origin.
 
\section{Conclusions}
\label{sec:conc}
For the first time, gravity wave-induced CO$_2$ density fluctuations in the
Martian upper thermosphere (180--220 km) observed by the NGIMS instrument
on board the NASA's MAVEN orbiter have been presented and discussed in the light
of direct propagation and dissipation of harmonics of lower atmospheric origin.
The main results and conclusions of this study are:
\begin{enumerate}
\item Evidence of wave-like perturbations is seen in the CO$_2$ number density
  observed by NGIMS during December 2014. The spatial scales of these
  perturbations, as measured along the MAVEN spacecraft
  trajectory, are consistent with the hypothesis that they are
  the manifestations of atmospheric gravity waves.
\item In the region studied here (180--220 km altitude,
  60$^\circ$--75$^\circ$ latitude and 2--10 hours local time), these waves have 
  typical amplitudes of 20--30\% relative to the background density.
\item An increase in the observed mean wave amplitude is seen with decreasing
  local time, which is similar to day-night variations observed by the ONMS mass
  spectrometer on Pioneer Venus. Some increase in the mean wave amplitude is
  seen as a function of latitude.
\item Numerical modeling with the extended gravity wave parameterization of
  \citet{Yigit_etal08} using the Mars Climate Database fields for the MAVEN
  spatiotemporal conditions shows, in qualitative agreement with the
  observations, that GW propagation into the thermosphere is highly variable and
  that the instantaneous density fluctuations are $>150 \%$ around 100 km and up
  to $50-100\%$ above 150 km.
\end{enumerate}

Overall, the wave-like density perturbations have been observed at all altitudes
up to 250 km. This is significantly higher than in previous observations of such
waves on Mars, and even higher than in the theoretical estimates using the GW
parameterization. More systematic measurements of density fluctuations in the
thermosphere are required to constrain GW parameterizations and elucidate the
physics and dynamical importance of these waves.

\begin{acknowledgments}
  Modeling data supporting Figures 3 and 4 are available from EY
  (eyigit@gmu.edu).
  Mars Climate Database is available at \\
  http://www-mars.lmd.jussieu.fr/mars/access.html.  This work was partially
  supported by NASA grant NNX13AO36G. The MAVEN project is supported by NASA
  through the Mars Exploration Program.
\end{acknowledgments}

\end{article}

\newpage
\end{document}